\documentclass[11pt]{article}
\usepackage{amssymb,amsmath,amsfonts}
\usepackage{graphicx}
\usepackage{graphics}
\usepackage{eepic,epsfig}

\begin{document}

\title{Wightman function and vacuum densities in de Sitter spacetime with
toroidally compactified dimensions}
\author{ S. Bellucci$^{1}$\thanks{%
E-mail: bellucci@lnf.infn.it } and A. A. Saharian$^{2}$\thanks{%
E-mail: saharian@ictp.it } \\
\textit{$^1$ INFN, Laboratori Nazionali di Frascati,}\\
\textit{Via Enrico Fermi 40,00044 Frascati, Italy} \\
\textit{$^2$ Department of Physics, Yerevan State University,}\\
\textit{1 Alex Manoogian Street, 0025 Yerevan, Armenia }}
\maketitle

\begin{abstract}
We investigate the Wightman function, the vacuum expectation values of the
field square and the energy-momentum tensor for a scalar field with general
curvature coupling parameter in $(D+1)$-dimensional de Sitter spacetime with
an arbitrary number of compactified spatial dimensions. Both cases of
periodicity and antiperiodicity conditions along the compactified dimensions
are considered. Recurrence formulae are derived which express the vacuum
expectation values for the dS spacetime of topology $\mathrm{R}^{p}\times (%
\mathrm{S}^{1})^{q}$ in the form of the sum of the vacuum expectation values
in the topology $\mathrm{R}^{p+1}\times (\mathrm{S}^{1})^{q-1}$ and the part
induced by the compactness of the $(p+1)$th spatial dimension. The behavior
of the topological parts is investigated in various asymptotic regions of
the parameters. In the early stages of the cosmological evolution the
topological parts dominate the contribution in the expectation values due to
the uncompactified dS part. In this limit the behavior of the topological
parts does not depend on the curvature coupling parameter and coincides with
that for a conformally coupled massless field. At late stages of the
cosmological expansion the expectation values are dominated by the part
corresponding to uncompactified dS spacetime. The vanishing of the
topological parts is monotonic or oscillatory in dependence of the mass and
the curvature coupling parameter of the field.
\end{abstract}

\bigskip

\section{Introduction}

De Sitter (dS) spacetime is among the most popular backgrounds in
gravitational physics. There are several reasons for this. First of all dS
spacetime is the maximally symmetric solution of Einstein's equation with a
positive cosmological constant. Due to the high symmetry numerous physical
problems are exactly solvable on this background. A better understanding of
physical effects in this background could serve as a handle to deal with
more complicated geometries. De Sitter spacetime plays an important role in
most inflationary models, where an approximately dS spacetime is employed to
solve a number of problems in standard cosmology \cite{Lind90}. More
recently astronomical observations of high redshift supernovae, galaxy
clusters and cosmic microwave background \cite{Ries07} indicate that at the
present epoch the universe is accelerating and can be well approximated by a
world with a positive cosmological constant. If the universe would
accelerate indefinitely, the standard cosmology would lead to an asymptotic
dS universe. In addition to the above, an interesting topic which has
received increasing attention is related to string-theoretical models of dS
spacetime and inflation. Recently a number of constructions of metastable dS
vacua within the framework of string theory are discussed (see, for
instance, \cite{Kach03,Silv07} and references therein).

There is no reason to believe that the version of dS spacetime which may
emerge from string theory, will necessarily be the most familiar version
with symmetry group $O(1,4)$ and there are many different topological spaces
which can accept the dS metric locally. There are many reasons to expect
that in string theory the most natural topology for the universe is that of
a flat compact three-manifold \cite{McIn04}. In particular, in Ref. \cite%
{Lind04} it was argued that from an inflationary point of view universes
with compact spatial dimensions, under certain conditions, should be
considered a rule rather than an exception. The models of a compact universe
with nontrivial topology may play an important role by providing proper
initial conditions for inflation (for the cosmological consequences of the
nontrivial topology and observational bounds on the size of compactified
dimensions see, for example, \cite{Lach95}). The quantum creation of the
universe having toroidal spatial topology is discussed in \cite{Zeld84} and
in references \cite{Gonc85} within the framework of various supergravity
theories. The compactification of spatial dimensions leads to the
modification of the spectrum of vacuum fluctuations and, as a result, to
Casimir-type contributions to the vacuum expectation values of physical
observables (for the topological Casimir effect and its role in cosmology
see \cite{Most97,Bord01,Eliz06} and references therein). The effect of the
compactification of a single spatial dimension in dS spacetime (topology $%
\mathrm{R}^{D-1}\times \mathrm{S}^{1}$) on the properties of quantum vacuum
for a scalar field with general curvature coupling parameter and with
periodicity condition along the compactified dimension is investigated in
Ref. \cite{Saha07} (for quantum effects in braneworld models with dS spaces
see, for instance, \cite{dSbrane}).

In view of the above mentioned importance of toroidally compactified dS
spacetimes, in the present paper we consider a general class of
compactifications having the spatial topology $\mathrm{R}^{p}\times (\mathrm{%
S}^{1})^{q}$, $p+q=D$. This geometry can be used to describe two types of
models. For the first one $p=3$, $q\geqslant 1$,\ and which corresponds to
the universe with Kaluza-Klein type extra dimensions. As it will be shown in
the present work, the presence of extra dimensions generates an additional
gravitational source in the cosmological equations which is of barotropic
type at late stages of the cosmological evolution. For the second model $D=3$
and the results given below describe how the properties of the universe with
dS geometry are changed by one-loop quantum effects induced by the
compactness of spatial dimensions. In quantum field theory on curved
backgrounds among the important quantities describing the local properties
of a quantum field and quantum back-reaction effects are the expectation
values of the field square and the energy-momentum tensor for a given
quantum state. In particular, the vacuum expectation values of these
quantities are of special interest. In order to evaluate these expectation
values, we construct firstly the corresponding positive frequency Wightman
function. Applying to the mode-sum the Abel-Plana summation formula, we
present this function as the sum of the Wightman function for the topology $%
\mathrm{R}^{p+1}\times (\mathrm{S}^{1})^{q-1}$ plus an additional term
induced by the compactness of the $(p+1)$th dimension. The latter is finite
in the coincidence limit and can be directly used for the evaluation of the
corresponding parts in the expectation \ values of the field square and the
energy-momentum tensor. In this way the renormalization of these quantities
is reduced to the renormalization of the corresponding quantities in
uncompactified dS spacetime. Note that for a scalar field on the background
of dS spacetime the renormalized vacuum expectation values of the field
square and the energy-momentum tensor are investigated in Refs. \cite%
{Cand75,Dowk76,Bunc78} by using various regularization schemes (see also
\cite{Birr82}). The corresponding effects upon phase transitions in an
expanding universe are discussed in \cite{Vile82,Alle83}.

The paper is organized as follows. In the next section we consider the
positive frequency Wightman function for dS spacetime of topology $\mathrm{R}%
^{p}\times (\mathrm{S}^{1})^{q}$. In sections \ref{sec:vevPhi2} and \ref%
{sec:vevEMT2} we use the formula for the Wightman function for the
evaluation of the vacuum expectation values of the field square and the
energy-momentum tensor. The asymptotic behavior of these quantities is
investigated in the early and late stages of the cosmological evolution. The
case of a twisted scalar field with antiperiodic boundary conditions is
considered in section \ref{sec:Twisted}. The main results of the paper are
summarized in section \ref{sec:Conc}.

\section{Wightman function in de Sitter spacetime with toroidally
compactified dimensions}

\label{sec:WF}

We consider a free massive scalar field with curvature coupling parameter $%
\xi $\ on background of $(D+1)$-dimensional de Sitter spacetime ($\mathrm{dS}%
_{D+1}$) generated by a positive cosmological constant $\Lambda $. The field
equation has the form%
\begin{equation}
\left( \nabla _{l}\nabla ^{l}+m^{2}+\xi R\right) \varphi =0,  \label{fieldeq}
\end{equation}%
where $R=2(D+1)\Lambda /(D-1)$ is the Ricci scalar for $\mathrm{dS}_{D+1}$
and $\xi $ is the curvature coupling parameter. The special cases $\xi =0$
and $\xi =\xi _{D}\equiv (D-1)/4D$ correspond to minimally and conformally
coupled fields respectively. The importance of these special cases is
related to that in the massless limit the corresponding fields mimic the
behavior of gravitons and photons. We write the line element for $\mathrm{dS}%
_{D+1}$ in planar (inflationary) coordinates most appropriate for
cosmological applications:%
\begin{equation}
ds^{2}=dt^{2}-e^{2t/\alpha }\sum_{i=1}^{D}(dz^{i})^{2},  \label{ds2deSit}
\end{equation}%
where the parameter $\alpha $ is related to the cosmological constant by the
formula%
\begin{equation}
\alpha ^{2}=\frac{D(D-1)}{2\Lambda }.  \label{alfa}
\end{equation}%
Below, in addition to the synchronous time coordinate $t$ we will also use
the conformal time $\tau $ in terms of which the line element takes
conformally flat form:%
\begin{equation}
ds^{2}=(\alpha /\tau )^{2}[d\tau ^{2}-\sum_{i=1}^{D}(dz^{i})^{2}],\;\tau
=-\alpha e^{-t/\alpha },\;-\infty <\tau <0.  \label{ds2Dd}
\end{equation}%
We assume that the spatial coordinates $z^{l}$, $l=p+1,\ldots ,D$, are
compactified to $\mathrm{S}^{1}$ of the length $L_{l}$: $0\leqslant
z^{l}\leqslant L_{l}$, and for the other coordinates we have $-\infty
<z^{l}<+\infty $, $l=1,\ldots ,p$. Hence, we consider the spatial topology $%
\mathrm{R}^{p}\times (\mathrm{S}^{1})^{q}$, where $q=D-p$. For $p=0$, as a
special case we obtain the toroidally compactified dS spacetime discussed in
\cite{McIn04,Lind04,Zeld84}. The Casimir densities for a scalar field with
periodicity conditions in the case $q=1$ were discussed previously in Ref.
\cite{Saha07}.

In the discussion below we will denote the position vectors along the
uncompactified and compactified dimensions by $\mathbf{z}_{p}=(z^{1},\ldots
,z^{p})$ and $\mathbf{z}_{q}=(z^{p+1},\ldots ,z^{D})$. For a scalar field
with periodic boundary condition one has (no summation over $l$)%
\begin{equation}
\varphi (t,\mathbf{z}_{p},\mathbf{z}_{q}+L_{l}\mathbf{e}_{l})=\varphi (t,%
\mathbf{z}_{p},\mathbf{z}_{q}),  \label{periodicBC}
\end{equation}%
where $l=p+1,\ldots ,D$ and $\mathbf{e}_{l}$ is the unit vector along the
direction of the coordinate $z^{l}$. In this paper we are interested in the
effects of non-trivial topology on the vacuum expectation values (VEVs) of
the field square and the energy-momentum tensor. These VEVs are obtained
from the corresponding positive frequency Wightman function $%
G_{p,q}^{+}(x,x^{\prime })$ in the coincidence limit of the arguments. The
Wightman function is also important in consideration of the response of
particle detectors at a given state of motion (see, for instance, \cite%
{Birr82}). Expanding the field operator over the complete set $\left\{
\varphi _{\sigma }(x),\varphi _{\sigma }^{\ast }(x)\right\} $ of positive
and negative frequency solutions to the classical field equation, satisfying
the periodicity conditions along the compactified dimensions, the positive
frequency Wightman function is presented as the mode-sum:
\begin{equation}
G_{p,q}^{+}(x,x^{\prime })=\langle 0|\varphi (x)\varphi (x^{\prime
})|0\rangle =\sum_{\sigma }\varphi _{\sigma }(x)\varphi _{\sigma }^{\ast
}(x^{\prime }),  \label{Wigh1}
\end{equation}%
where the collective index $\sigma $ specifies the solutions.

Due to the symmetry of the problem under consideration the spatial
dependence of the eigenfunctions $\varphi _{\sigma }(x)$ can be taken in the
standard plane-wave form, $e^{i\mathbf{k}\cdot \mathbf{z}}$. Substituting
into the field equation, we obtain that the time dependent part of the
eigenfunctions is a linear combination of the functions $\tau ^{D/2}H_{\nu
}^{(l)}(|\mathbf{k|}\tau )$, $l=1,2$, where $H_{\nu }^{(l)}(x)$ is the
Hankel function and
\begin{equation}
\nu =\left[ D^{2}/4-D(D+1)\xi -m^{2}\alpha ^{2}\right] ^{1/2}.  \label{knD}
\end{equation}%
Different choices of the coefficients in this linear combination correspond
to different choices of the vacuum state. We will consider de Sitter
invariant Bunch-Davies vacuum \cite{Bunc78} for which the coefficient for
the part containing the function $H_{\nu }^{(1)}(|\mathbf{k|}\tau )$ is
zero. The corresponding eigenfunctions satisfying the periodicity conditions
take the form
\begin{equation}
\varphi _{\sigma }(x)=C_{\sigma }\eta ^{D/2}H_{\nu }^{(1)}(k\eta )e^{i%
\mathbf{k}_{p}\cdot \mathbf{z}_{p}+i\mathbf{k}_{q}\cdot \mathbf{z}%
_{q}},\;\eta =\alpha e^{-t/\alpha },  \label{eigfuncD}
\end{equation}%
where we have decomposed the contributions from the uncompactified and
compactified dimensions with the notations%
\begin{eqnarray}
\mathbf{k}_{p} &=&(k_{1},\ldots ,k_{p}),\;\mathbf{k}_{q}=(k_{p+1},\ldots
,k_{D}),\;k=\sqrt{\mathbf{k}_{p}^{2}+\mathbf{k}_{q}^{2}},\;  \notag \\
\;k_{l} &=&2\pi n_{l}/L_{l},\;n_{l}=0,\pm 1,\pm 2,\ldots ,\;l=p+1,\ldots ,D.
\label{kD1D2}
\end{eqnarray}%
Note that we have transformed the Hankel function to have the positive
defined argument and instead of the conformal time $\tau $ the variable $%
\eta $ is introduced which we will call the conformal time as well. The
eigenfunctions are specified by the set $\sigma =(\mathbf{k}%
_{p},n_{p+1},\ldots ,n_{D})$ and the coefficient $C_{\sigma }$ is found from
the standard orthonormalization condition
\begin{equation}
-i\int d^{D}x\sqrt{|g|}g^{00}\varphi _{\sigma }(x)\overleftrightarrow{%
\partial }_{\tau }\varphi _{\sigma ^{\prime }}^{\ast }(x)=\delta _{\sigma
\sigma ^{\prime }},  \label{normcond}
\end{equation}%
where the integration goes over the spatial hypersurface $\tau =\mathrm{const%
}$, and $\delta _{\sigma \sigma ^{\prime }}$ is understood as the Kronecker
delta for the discrete indices and as the Dirac delta-function for the
continuous ones. By using the Wronskian relation for the Hankel functions
one finds%
\begin{equation}
C_{\sigma }^{2}=\frac{\alpha ^{1-D}e^{i(\nu -\nu ^{\ast })\pi /2}}{%
2^{p+2}\pi ^{p-1}L_{p+1}\cdots L_{D}}.  \label{normCD}
\end{equation}

Having the complete set of eigenfunctions and using the mode-sum formula (%
\ref{Wigh1}), for the positive frequency Wightman function we obtain the
formula
\begin{eqnarray}
G_{p,q}^{+}(x,x^{\prime }) &=&\frac{\alpha ^{1-D}(\eta \eta ^{\prime
})^{D/2}e^{i(\nu -\nu ^{\ast })\pi /2}}{2^{p+2}\pi ^{p-1}L_{p+1}\cdots L_{D}}%
\int d\mathbf{k}_{p}\,e^{i\mathbf{k}_{p}\cdot \Delta \mathbf{z}_{p}}  \notag
\\
&&\times \sum_{\mathbf{n}_{q}=-\infty }^{+\infty }e^{i\mathbf{k}_{q}\cdot
\Delta \mathbf{z}_{q}}H_{\nu }^{(1)}(k\eta )[H_{\nu }^{(1)}(k\eta ^{\prime
})]^{\ast },  \label{GxxD}
\end{eqnarray}%
with $\Delta \mathbf{z}_{p}=\mathbf{z}_{p}-\mathbf{z}_{p}^{\prime }$, $%
\Delta \mathbf{z}_{q}=\mathbf{z}_{q}-\mathbf{z}_{q}^{\prime }$, and%
\begin{equation}
\sum_{\mathbf{n}_{q}=-\infty }^{+\infty }=\sum_{n_{p+1}=-\infty }^{+\infty
}\ldots \sum_{n_{D}=-\infty }^{+\infty }.  \label{nqsum}
\end{equation}%
As a next step, we apply to the series over $n_{p+1}$ in (\ref{GxxD}) the
Abel-Plana formula \cite{Most97,Saha07Gen}%
\begin{equation}
\sideset{}{'}{\sum}_{n=0}^{\infty }f(n)=\int_{0}^{\infty
}dx\,f(x)+i\int_{0}^{\infty }dx\,\frac{f(ix)-f(-ix)}{e^{2\pi x}-1},
\label{Abel}
\end{equation}%
where the prime means that the term $n=0$ should be halved. It can be seen
that after the application of this formula the term in the expression of the
Wightman function which corresponds to the first integral on the right of (%
\ref{Abel}) is the Wightman function for dS spacetime with the topology $%
\mathrm{R}^{p+1}\times (\mathrm{S}^{1})^{q-1}$, which, in the notations
given above, corresponds to the function $G_{p+1,q-1}^{+}(x,x^{\prime })$.
As a result one finds
\begin{equation}
G_{p,q}^{+}(x,x^{\prime })=G_{p+1,q-1}^{+}(x,x^{\prime })+\Delta
_{p+1}G_{p,q}^{+}(x,x^{\prime }).  \label{G1decomp}
\end{equation}%
The second term on the right of this formula is induced by the compactness
of the $z^{p+1}$ - direction and is given by the expression
\begin{eqnarray}
\Delta _{p+1}G_{p,q}^{+}(x,x^{\prime }) &=&\frac{2\alpha ^{1-D}(\eta \eta
^{\prime })^{D/2}}{(2\pi )^{p+1}V_{q-1}}\int d\mathbf{k}_{p}\,e^{i\mathbf{k}%
_{p}\cdot \Delta \mathbf{z}_{p}}\sum_{\mathbf{n}_{q-1}=-\infty }^{+\infty
}e^{i\mathbf{k}_{q-1}\cdot \Delta \mathbf{z}_{q-1}}  \notag \\
&&\times \int_{0}^{\infty }dx\,\frac{x\cosh (\sqrt{x^{2}+\mathbf{k}%
_{p}^{2}+k_{\mathbf{n}_{q-1}}^{2}}\Delta z^{p+1})}{\sqrt{x^{2}+\mathbf{k}%
_{p}^{2}+k_{\mathbf{n}_{q-1}}^{2}}(e^{L_{p+1}\sqrt{x^{2}+\mathbf{k}%
_{p}^{2}+k_{\mathbf{n}_{q-1}}^{2}}}-1)}  \notag \\
&&\times \left[ K_{\nu }(\eta x)I_{-\nu }(\eta ^{\prime }x)+I_{\nu }(\eta
x)K_{\nu }(\eta ^{\prime }x)\right] ,  \label{GxxD2}
\end{eqnarray}%
where $\mathbf{n}_{q-1}=(n_{p+2},\ldots ,n_{D})$, $I_{\nu }(x)$ and $K_{\nu
}(x)$ are the Bessel modified functions and the notation%
\begin{equation}
k_{\mathbf{n}_{q-1}}^{2}=\sum_{l=p+2}^{D}(2\pi n_{l}/L_{l})^{2}
\label{knD1+2}
\end{equation}%
is introduced. In formula (\ref{GxxD2}), $V_{q-1}=L_{p+2}\cdots L_{D}$ is
the volume of $(q-1)$-dimensional compact subspace. Note that the
combination of the Bessel modified functions appearing in formula (\ref%
{GxxD2}) can also be written in the form%
\begin{eqnarray}
K_{\nu }(\eta x)I_{-\nu }(\eta ^{\prime }x)+I_{\nu }(\eta x)K_{\nu }(\eta
^{\prime }x) &=&\frac{2}{\pi }\sin (\nu \pi )K_{\nu }(\eta x)K_{\nu }(\eta
^{\prime }x)  \notag \\
&&+I_{\nu }(\eta x)K_{\nu }(\eta ^{\prime }x)+K_{\nu }(\eta x)I_{\nu }(\eta
^{\prime }x),  \label{eqformComb}
\end{eqnarray}%
which explicitly shows that this combination is symmetric under the
replacement $\eta \rightleftarrows \eta ^{\prime }$. In formula (\ref{GxxD2}%
) the integration with respect to the angular part of $\mathbf{k}_{p}$ can
be done by using the formula%
\begin{equation}
\int d\mathbf{k}_{p}\,e^{i\mathbf{k}_{p}\cdot \Delta \mathbf{z}_{p}}F(|%
\mathbf{k}_{p}|)=\frac{(2\pi )^{p/2}}{|\Delta \mathbf{z}_{p}|^{p/2-1}}%
\int_{0}^{\infty }d|\mathbf{k}_{p}|\,|\mathbf{k}_{p}|^{p/2}F(|\mathbf{k}%
_{p}|)J_{p/2-1}(|\mathbf{k}_{p}||\Delta \mathbf{z}_{p}|),  \label{intang}
\end{equation}%
where $J_{\mu }(x)$ is the Bessel function.

After the recurring application of formula (\ref{GxxD2}), the Wightman
function for dS spacetime with spatial topology $\mathrm{R}^{p}\times (%
\mathrm{S}^{1})^{q}$ is presented in the form%
\begin{equation}
G_{p,q}^{+}(x,x^{\prime })=G_{\mathrm{dS}}^{+}(x,x^{\prime })+\Delta
G_{p,q}^{+}(x,x^{\prime }),  \label{GdSGcomp}
\end{equation}%
where $G_{\mathrm{dS}}^{+}(x,x^{\prime })\equiv G_{D,0}^{+}(x,x^{\prime })$
is the corresponding function for uncompactified dS spacetime and the part%
\begin{equation}
\Delta G_{p,q}^{+}(x,x^{\prime })=\sum_{l=1}^{q}\Delta
_{D-l+1}G_{D-l,l}^{+}(x,x^{\prime }),  \label{DeltaGtop}
\end{equation}%
is induced by the toroidal compactification of the $q$-dimensional subspace.
Two-point function in the uncompactified dS spacetime is investigated in
\cite{Cand75,Dowk76,Bunc78,Bros96,Bous02} (see also \cite{Birr82}) and is
given by the formula%
\begin{equation}
G_{\mathrm{dS}}^{+}(x,x^{\prime })=\frac{\alpha ^{1-D}\Gamma (D/2+\nu
)\Gamma (D/2-\nu )}{2^{(D+3)/2}\pi ^{(D+1)/2}\left( u^{2}-1\right) ^{(D-1)/4}%
}P_{\nu -1/2}^{(1-D)/2}(u),  \label{WFdS}
\end{equation}%
where $P_{\nu }^{\mu }(x)$ is the associated Legendre function of the first
kind and
\begin{equation}
u=-1+\frac{\sum_{l=1}^{D}(z^{l}-z^{\prime l})^{2}-(\eta -\eta ^{\prime })^{2}%
}{2\eta \eta ^{\prime }}.  \label{u}
\end{equation}%
An alternative form is obtained by using the relation between the the
associated Legendre function and the hypergeometric function.

\section{Vacuum expectation values of the field square}

\label{sec:vevPhi2}

We denote by $\langle \varphi ^{2}\rangle _{p,q}$ the VEV of the field
square in dS spacetime with spatial topology $\mathrm{R}^{p}\times (\mathrm{S%
}^{1})^{q}$. Having the Wightman function we can evaluate this VEV taking
the coincidence limit of the arguments. Of course, in this limit the
two-point functions are divergent and some renormalization procedure is
needed. The important point here is that the local geometry is not changed
by the toroidal compactification and the divergences are the same as in the
uncompactified dS spacetime. As in our procedure we have already extracted
from the Wightman function the part $G_{\mathrm{dS}}^{+}(x,x^{\prime })$,
the renormalization of the VEVs is reduced to the renormalization of the
uncompactified dS part which is already done in literature. The VEV\ of the
field square is presented in the decomposed form%
\begin{equation}
\langle \varphi ^{2}\rangle _{p,q}=\langle \varphi ^{2}\rangle _{\mathrm{dS}%
}+\langle \varphi ^{2}\rangle _{c},\;\langle \varphi ^{2}\rangle
_{c}=\sum_{l=1}^{q}\Delta _{D-l+1}\langle \varphi ^{2}\rangle _{D-l,l},
\label{phi2dSplComp}
\end{equation}%
where $\langle \varphi ^{2}\rangle _{\mathrm{dS}}$ is the VEV in
uncompactified $\mathrm{dS}_{D+1}$ and the part $\langle \varphi ^{2}\rangle
_{c}$ is due to the compactness of the $q$-dimensional subspace. Here the
term $\Delta _{p+1}\langle \varphi ^{2}\rangle _{p,q}$ is defined by the
relation similar to (\ref{G1decomp}):
\begin{equation}
\langle \varphi ^{2}\rangle _{p,q}=\langle \varphi ^{2}\rangle
_{p+1,q-1}+\Delta _{p+1}\langle \varphi ^{2}\rangle _{p,q}.
\label{phi2decomp}
\end{equation}%
This term is the part in the VEV induced by the compactness of the $z^{p+1}$
- direction. This part is directly obtained from (\ref{GxxD2}) in the
coincidence limit of the arguments:%
\begin{eqnarray}
\Delta _{p+1}\langle \varphi ^{2}\rangle _{p,q} &=&\frac{2\alpha ^{1-D}\eta
^{D}}{2^{p}\pi ^{p/2+1}\Gamma (p/2)V_{q-1}}\sum_{\mathbf{n}_{q-1}=-\infty
}^{+\infty }\int_{0}^{\infty }d|\mathbf{k}_{p}|\,|\mathbf{k}_{p}|^{p-1}
\notag \\
&&\times \int_{0}^{\infty }dx\,\frac{xK_{\nu }(x\eta )\left[ I_{-\nu }(x\eta
)+I_{\nu }(x\eta )\right] }{\sqrt{x^{2}+\mathbf{k}_{p}^{2}+k_{\mathbf{n}%
_{q-1}}^{2}}(e^{L_{p+1}\sqrt{x^{2}+\mathbf{k}_{p}^{2}+k_{\mathbf{n}%
_{q-1}}^{2}}}-1)}.  \label{phi2Dc}
\end{eqnarray}%
Instead of $|\mathbf{k}_{p}|$ introducing a new integration variable $y=%
\sqrt{x^{2}+\mathbf{k}_{p}^{2}+k_{\mathbf{n}_{q-1}}^{2}}$ and expanding $%
(e^{Ly}-1)^{-1}$, the integral over $y$ is explicitly evaluated and one finds%
\begin{eqnarray}
\Delta _{p+1}\langle \varphi ^{2}\rangle _{p,q} &=&\frac{4\alpha ^{1-D}\eta
^{D}}{(2\pi )^{(p+3)/2}V_{q-1}}\sum_{n=1}^{\infty }\sum_{\mathbf{n}%
_{q-1}=-\infty }^{+\infty }\int_{0}^{\infty }dx\,xK_{\nu }(x\eta )  \notag \\
&&\times \frac{I_{-\nu }(x\eta )+I_{\nu }(x\eta )}{(nL_{p+1})^{p-1}}%
f_{(p-1)/2}(nL_{p+1}\sqrt{x^{2}+k_{\mathbf{n}_{q-1}}^{2}}),  \label{DelPhi2}
\end{eqnarray}%
where we use the notation%
\begin{equation}
f_{\mu }(y)=y^{\mu }K_{\mu }(y).  \label{fmunot}
\end{equation}%
By taking into account the relation between the conformal and synchronous
time coordinates, we see that the VEV of the field square is a function of
the combinations $L_{l}/\eta =L_{l}e^{t/\alpha }/\alpha $. In the limit when
the length of the one of the compactified dimensions, say $z^{l}$, $%
l\geqslant p+2$, is large, $L_{l}\rightarrow \infty $, the main contribution
into the sum over $n_{l}$ in (\ref{DelPhi2}) comes from large values of $%
n_{l}$ and we can replace the summation by the integration in accordance
with the formula%
\begin{equation}
\frac{1}{L_{l}}\sum_{n_{l}=-\infty }^{+\infty }f(2\pi n_{l}/L_{l})=\frac{1}{%
\pi }\int_{0}^{\infty }dy\,f(y).  \label{sumtoint}
\end{equation}%
The integral over $y$ is evaluated by using the formula from \cite{Prud86}
and we can see that from (\ref{DelPhi2}) the corresponding formula is
obtained for the topology $\mathrm{R}^{p+1}\times (\mathrm{S}^{1})^{q-1}$.

For a conformally coupled massless scalar field one has $\nu =1/2$ and $%
\left[ I_{-\nu }(x)+I_{\nu }(x)\right] K_{\nu }(x)=1/x$. In this case the
corresponding integral in formula (\ref{DelPhi2}) is explicitly evaluated
and we find%
\begin{equation}
\Delta _{p+1}\langle \varphi ^{2}\rangle _{p,q}=\frac{2(\eta /\alpha )^{D-1}%
}{(2\pi )^{p/2+1}V_{q-1}}\sum_{n=1}^{\infty }\sum_{\mathbf{n}_{q-1}=-\infty
}^{+\infty }\frac{f_{p/2}(nL_{p+1}k_{\mathbf{n}_{q-1}})}{(L_{p+1}n)^{p}}%
,\;\xi =\xi _{D},\;m=0.  \label{DelPhi2Conf}
\end{equation}%
In particular, the topological part is always positive. Formula (\ref%
{DelPhi2Conf}) could also be obtained from the corresponding result in $%
(D+1) $-dimensional Minkowski spacetime with spatial topology $\mathrm{R}%
^{p}\times (\mathrm{S}^{1})^{q}$, taking into account that two problems are
conformally related: $\Delta _{p+1}\langle \varphi ^{2}\rangle
_{p,q}=a^{1-D}(\eta )\Delta _{p+1}\langle \varphi ^{2}\rangle _{p,q}^{%
\mathrm{(M)}}$, where $a(\eta )=\alpha /\eta $ is the scale factor. This
relation is valid for any conformally flat bulk. The similar formula takes
place for the total topological part $\langle \varphi ^{2}\rangle _{c}$.
Note that, in this case the expressions for $\Delta _{p+1}\langle \varphi
^{2}\rangle _{p,q}$ are obtained from the formulae for $\Delta _{p+1}\langle
\varphi ^{2}\rangle _{p,q}^{\mathrm{(M)}}$ replacing the lengths $L_{l}$ of
the compactified dimensions by the comoving lengths $\alpha L_{l}/\eta $, $%
l=p,\ldots ,D$.

Now we turn to the investigation of the topological part $\Delta
_{p+1}\langle \varphi ^{2}\rangle _{p,q}$ in the VEV of the field square in
the asymptotic regions of the ratio $L_{p+1}/\eta $. For small values of
this ratio, $L_{p+1}/\eta \ll 1$, we introduce a new integration variable $%
y=L_{p+1}x$. By taking into account that for large values $x$ one has $\left[
I_{-\nu }(x)+I_{\nu }(x)\right] K_{\nu }(x)\approx 1/x$, we find that to the
leading order $\Delta _{p+1}\langle \varphi ^{2}\rangle _{p,q}$ coincides
with the corresponding result for a conformally coupled massless field,
given by (\ref{DelPhi2Conf}):%
\begin{equation}
\Delta _{p+1}\langle \varphi ^{2}\rangle _{p,q}\approx (\eta /\alpha
)^{D-1}\Delta _{p+1}\langle \varphi ^{2}\rangle _{p,q}^{\mathrm{(M)}%
},\;L_{p+1}/\eta \ll 1.  \label{DelPhi2Poq}
\end{equation}%
For fixed value of the ratio $L_{p+1}/\alpha $, this limit corresponds to $%
t\rightarrow -\infty $ and the topological part $\langle \varphi ^{2}\rangle
_{c}$ behaves like $\exp [-(D-1)t/\alpha ]$. By taking into account that the
part $\langle \varphi ^{2}\rangle _{\mathrm{dS}}$ is time independent, from
here we conclude that in the early stages of the cosmological expansion the
topological part dominates in the VEV\ of the field square.

For small values of the ratio $\eta /L_{p+1}$, we introduce a new
integration variable $y=L_{p+1}x$ and expand the integrand by using the
formulae for the Bessel modified functions for small arguments. For real
values of the parameter $\nu $, after the integration over $y$ by using the
formula from \cite{Prud86}, to the leading order we find%
\begin{equation}
\Delta _{p+1}\langle \varphi ^{2}\rangle _{p,q}\approx \frac{2^{(1-p)/2+\nu
}\eta ^{D-2\nu }\Gamma (\nu )}{\pi ^{(p+3)/2}V_{q-1}\alpha ^{D-1}}%
\sum_{n=1}^{\infty }\sum_{\mathbf{n}_{q-1}=-\infty }^{+\infty }\frac{%
f_{(p+1)/2-\nu }(nL_{p+1}k_{\mathbf{n}_{q-1}})}{(L_{p+1}n)^{p+1-2\nu }}%
,\;\eta /L_{p+1}\ll 1.  \label{DelPhi2Mets}
\end{equation}%
In the case of a conformally coupled massless scalar field $\nu =1/2$ and
this formula reduces to the exact result given by Eq. (\ref{DelPhi2Conf}).
For fixed values of $L_{p+1}/\alpha $, the limit under consideration
corresponds to late stages of the cosmological evolution, $t\rightarrow
+\infty $, and the topological part $\langle \varphi ^{2}\rangle _{c}$ is
suppressed by the factor $\exp [-(D-2\nu )t/\alpha ]$. Hence, in this limit
the total VEV is dominated by the uncompactified dS part $\langle \varphi
^{2}\rangle _{\mathrm{dS}}$. Note that formula (32) also describes the
asymptotic behavior of the topological part in the strong curvature regime
corresponding to small values of the parameter $\alpha $.

In the same limit, for pure imaginary values of the parameter $\nu $ in a
similar way we find the following asymptotic behavior

\begin{eqnarray}
\Delta _{p+1}\langle \varphi ^{2}\rangle _{p,q} &\approx &\frac{4\alpha
^{1-D}\eta ^{D}}{(2\pi )^{(p+3)/2}V_{q-1}}\sum_{n=1}^{\infty }\sum_{\mathbf{n%
}_{q-1}=-\infty }^{+\infty }\frac{1}{(nL_{p+1})^{p+1}}  \notag \\
&&\times {\mathrm{Re}}\left[ 2^{i|\nu |}\Gamma (i|\nu |)(nL_{p+1}/\eta
)^{2i|\nu |}f_{(p+1)/2-i|\nu |}(nL_{p+1}k_{\mathbf{n}_{q-1}})\right] .
\label{DelPhi2MetsIm}
\end{eqnarray}%
Defining the phase $\phi _{0}$ by the relation

\begin{equation}
Be^{i\phi _{0}}=2^{i|\nu |}\Gamma (i|\nu |)\sum_{n=1}^{\infty }\sum_{\mathbf{%
n}_{q-1}=-\infty }^{+\infty }n^{2i|\nu |-p-1}f_{(p+1)/2-i|\nu |}(nL_{p+1}k_{%
\mathbf{n}_{q-1}}),  \label{Bphi0}
\end{equation}%
we write this formula in terms of the synchronous time:%
\begin{equation}
\Delta _{p+1}\langle \varphi ^{2}\rangle _{p,q}\approx \frac{4\alpha
e^{-Dt/\alpha }B}{(2\pi )^{(p+3)/2}L_{p+1}^{p+1}V_{q-1}}\cos [2|\nu
|t/\alpha +2|\nu |\ln (L_{p+1}/\alpha )+\phi _{0}].  \label{DelPhi2MetsIm1}
\end{equation}%
Hence, in the case under consideration at late stages of the cosmological
evolution the topological part is suppressed by the factor $\exp (-Dt/\alpha
)$ and the damping of the corresponding VEV has an oscillatory nature.

\section{Vacuum energy-momentum tensor}

\label{sec:vevEMT2}

In this section we investigate the VEV for the energy-momentum tensor of a
scalar field in $\mathrm{dS}_{D+1}$ with toroidally compactified $q$%
-dimensional subspace. In addition to describing the physical structure of
the quantum field at a given point, this quantity acts as the source of
gravity in the semiclassical Einstein equations. It therefore plays an
important role in modelling self-consistent dynamics involving the
gravitational field. Having the Wightman function and the VEV of the field
square we can evaluate the vacuum energy-momentum tensor by using the formula%
\begin{equation}
\langle T_{ik}\rangle _{p,q}=\lim_{x^{\prime }\rightarrow x}\partial
_{i}\partial _{k}^{\prime }G_{p,q}^{+}(x,x^{\prime })+\left[ \left( \xi -%
\frac{1}{4}\right) g_{ik}\nabla _{l}\nabla ^{l}-\xi \nabla _{i}\nabla
_{k}-\xi R_{ik}\right] \langle \varphi ^{2}\rangle _{p,q},  \label{emtvev1}
\end{equation}%
where $R_{ik}=Dg_{ik}/\alpha ^{2}$ is the Ricci tensor for $\mathrm{dS}_{D+1}
$. Note that in (\ref{emtvev1}) we have used the expression for the
classical energy-momentum tensor which differs from the standard one by the
term which vanishes on the solutions of the field equation (see, for
instance, Ref. \cite{Saha04}). As in the case of the field square, the VEV
of the energy-momentum tensor is presented in the form%
\begin{equation}
\langle T_{i}^{k}\rangle _{p,q}=\langle T_{i}^{k}\rangle _{p+1,q-1}+\Delta
_{p+1}\langle T_{i}^{k}\rangle _{p,q}.  \label{TikDecomp}
\end{equation}%
Here $\langle T_{i}^{k}\rangle _{p+1,q-1}$ is the part corresponding to dS
spacetime with $p+1$ uncompactified and $q-1$ toroidally compactified
dimensions and $\Delta _{p+1}\langle T_{i}^{k}\rangle _{p,q}$ is induced by
the compactness along the $z^{p+1}$ - direction. The recurring application
of formula (\ref{TikDecomp}) allows us to write the VEV in the form%
\begin{equation}
\langle T_{i}^{k}\rangle _{p,q}=\langle T_{i}^{k}\rangle _{\mathrm{dS}%
}+\langle T_{i}^{k}\rangle _{c},\;\langle T_{i}^{k}\rangle
_{c}=\sum_{l=1}^{q}\Delta _{D-l+1}\langle T_{i}^{k}\rangle _{D-l,l},
\label{TikComp}
\end{equation}%
where the part corresponding to uncompactified dS spacetime, $\langle
T_{i}^{k}\rangle _{\mathrm{dS}}$, is explicitly decomposed. The part $%
\langle T_{i}^{k}\rangle _{c}$ is induced by the comactness of the $q$%
-dimensional subspace.

The second term on the right of formula (\ref{TikDecomp}) is obtained
substituting the corresponding parts in the Wightman function, Eq. (\ref%
{GxxD2}), and in the field square, Eq. (\ref{DelPhi2}), into formula (\ref%
{emtvev1}). After the lengthy calculations for the energy density one finds%
\begin{eqnarray}
\Delta _{p+1}\langle T_{0}^{0}\rangle _{p,q} &=&\frac{2\alpha ^{-1-D}\eta
^{D}}{(2\pi )^{(p+3)/2}V_{q-1}}\sum_{n=1}^{\infty }\sum_{\mathbf{n}%
_{q-1}=-\infty }^{+\infty }\int_{0}^{\infty }dx  \notag \\
&&\times \frac{xF^{(0)}(x\eta )}{(nL_{p+1})^{p-1}}f_{(p-1)/2}(nL_{p+1}\sqrt{%
x^{2}+k_{\mathbf{n}_{q-1}}^{2}}),  \label{DelT00}
\end{eqnarray}%
with the notation%
\begin{eqnarray}
F^{(0)}(y) &=&y^{2}\left[ I_{-\nu }^{\prime }(y)+I_{\nu }^{\prime }(y)\right]
K_{\nu }^{\prime }(y)+D(1/2-2\xi )y\left[ (I_{-\nu }(y)+I_{\nu }(y))K_{\nu
}(y)\right] ^{\prime }  \notag \\
&&+\left[ I_{-\nu }(y)+I_{\nu }(y)\right] K_{\nu }(y)\left( \nu
^{2}+2m^{2}\alpha ^{2}-y^{2}\right) ,  \label{F0}
\end{eqnarray}%
and the function $f_{\mu }(y)$ is defined by formula (\ref{fmunot}). The
vacuum stresses are presented in the form (no summation over $i$)%
\begin{eqnarray}
\Delta _{p+1}\langle T_{i}^{i}\rangle _{p,q} &=&A_{p,q}-\frac{4\alpha
^{-1-D}\eta ^{D+2}}{(2\pi )^{(p+3)/2}V_{q-1}}\sum_{n=1}^{\infty }\sum_{%
\mathbf{n}_{q-1}=-\infty }^{+\infty }\int_{0}^{\infty }dx\,xK_{\nu }(x\eta )
\notag \\
&&\times \frac{I_{-\nu }(x\eta )+I_{\nu }(x\eta )}{(nL_{p+1})^{p+1}}%
f_{p}^{(i)}(nL_{p+1}\sqrt{x^{2}+k_{\mathbf{n}_{q-1}}^{2}}),  \label{DelTii}
\end{eqnarray}%
where we have introduced the notations%
\begin{eqnarray}
f_{p}^{(i)}(y) &=&f_{(p+1)/2}(y),\;i=1,\ldots ,p,  \notag \\
f_{p}^{(p+1)}(y) &=&-y^{2}f_{(p-1)/2}(y)-pf_{(p+1)/2}(y),  \label{fp+1} \\
f_{p}^{(i)}(y) &=&(nL_{p+1}k_{i})^{2}f_{(p-1)/2}(y),\;i=p+2,\ldots ,D.
\notag
\end{eqnarray}%
In formula (\ref{DelTii}) (no summation over $i$, $i=1,\ldots ,D$),
\begin{eqnarray}
A_{p,q} &=&\left[ \left( \xi -\frac{1}{4}\right) \nabla _{l}\nabla ^{l}-\xi
g^{ii}\nabla _{i}\nabla _{i}-\xi R_{i}^{i}\right] \Delta _{p+1}\langle
\varphi ^{2}\rangle _{p,q}  \notag \\
&=&\frac{2\alpha ^{-1-D}\eta ^{D}}{(2\pi )^{(p+3)/2}V_{q-1}}%
\sum_{n=1}^{\infty }\sum_{\mathbf{n}_{q-1}=-\infty }^{+\infty
}\int_{0}^{\infty }dx\,\frac{xF(x\eta )}{(nL_{p+1})^{p-1}}%
f_{(p-1)/2}(nL_{p+1}\sqrt{x^{2}+k_{\mathbf{n}_{q-1}}^{2}}),  \label{A}
\end{eqnarray}%
with the notation%
\begin{eqnarray}
F(y) &=&\left( 4\xi -1\right) y^{2}\left[ I_{-\nu }^{\prime }(y)+I_{\nu
}^{\prime }(y)\right] K_{\nu }^{\prime }(y)+\left[ 2(D+1)\xi -D/2\right] y(%
\left[ I(y)+I_{\nu }(y)\right] K_{\nu }(y))^{\prime }  \notag \\
&&+\left[ I_{-\nu }(y)+I_{\nu }(y)\right] K_{\nu }(y)\left[ \left( 4\xi
-1\right) \left( y^{2}+\nu ^{2}\right) \right] .  \label{Fy}
\end{eqnarray}%
As it is seen from the obtained formulae, the topological parts in the VEVs
are time-dependent and, hence, the local dS symmetry is broken by them.

As an additional check of our calculations it can be seen that the
topological terms satisfy the trace relation
\begin{equation}
\Delta _{p+1}\langle T_{i}^{i}\rangle _{p,q}=D(\xi -\xi _{D})\nabla
_{l}\nabla ^{l}\Delta _{p+1}\langle \varphi ^{2}\rangle _{p,q}+m^{2}\Delta
_{p+1}\langle \varphi ^{2}\rangle _{p,q}.  \label{tracerel}
\end{equation}%
In particular, from here it follows that the topological part in the VEV\ of
the energy-momentum tensor is traceless for a conformally coupled massless
scalar field. The trace anomaly is contained in the uncompactified dS part
only. We could expect this result, as the trace anomaly is determined by the
local geometry and the local geometry is not changed by the toroidal
compactification.

For a conformally coupled massless scalar field $\nu =1/2$ and, by using the
formulae for $I_{\pm 1/2}(x)$ and $K_{1/2}(x)$, after the integration over $%
x $ from formulae (\ref{DelT00}), (\ref{DelTii}) we find (no summation over $%
i$)%
\begin{equation}
\Delta _{p+1}\langle T_{i}^{i}\rangle _{p,q}=-\frac{2(\eta /\alpha )^{D+1}}{%
(2\pi )^{p/2+1}V_{q-1}}\sum_{n=1}^{\infty }\sum_{\mathbf{n}_{q-1}=-\infty
}^{+\infty }\frac{g_{p}^{(i)}(nL_{p+1}k_{\mathbf{n}_{q-1}})}{(nL_{p+1})^{p+2}%
},  \label{DelTConf}
\end{equation}%
with the notations%
\begin{eqnarray}
g_{p}^{(0)}(y) &=&g_{p}^{(i)}(y)=f_{p/2+1}(y),\;i=1,\ldots ,p,  \notag \\
g_{p}^{(i)}(y) &=&(nL_{p+1}k_{i})^{2}f_{p/2}(y),\;i=p+2,\ldots ,D,
\label{gi} \\
g_{p}^{(p+1)}(y) &=&-(p+1)f_{p/2+1}(y)-y^{2}f_{p/2}(y).  \notag
\end{eqnarray}%
As in the case of the filed square, this formula can be directly obtained by
using the conformal relation between the problem under consideration and the
corresponding problem in $(D+1)$-dimensional Minkowski spacetime with the
spatial topology $\mathrm{R}^{p}\times (\mathrm{S}^{1})^{q}$. Note that in
this case the topological part in the energy density is always negative and
is equal to the vacuum stresses along the uncompactified dimensions. In
particular, for the case $D=3$, $p=0$ (topology $(\mathrm{S}^{1})^{3}$) and
for $L_{i}=L$, $i=1,2,3$, from formulae (\ref{TikComp}), (\ref{DelTConf})
for the topological part in the vacuum energy density we find $\langle
T_{0}^{0}\rangle _{c}=-0.8375(a(\eta )L)^{-4}$ (see, for example, Ref. \cite%
{Most97}).

The general formulae for the topological part in the VEV of the energy
density are simplified in the asymptotic regions of the parameters. For
small values of the ratio $L_{p+1}/\eta $ we can see that to the leading
order $\Delta _{p+1}\langle T_{i}^{k}\rangle _{p,q}$ coincides with the
corresponding result for a conformally coupled massless field (no summation
over $i$):%
\begin{equation}
\Delta _{p+1}\langle T_{i}^{i}\rangle _{p,q}\approx -\frac{2(\eta /\alpha
)^{D+1}}{(2\pi )^{p/2+1}V_{q-1}}\sum_{n=1}^{\infty }\sum_{\mathbf{n}%
_{q-1}=-\infty }^{+\infty }\frac{g_{p}^{(i)}(nL_{p+1}k_{\mathbf{n}_{q-1}})}{%
(nL_{p+1})^{p+2}},\;L/\eta \ll 1.  \label{TiiSmall}
\end{equation}%
For fixed values of the ratio $L_{p+1}/\alpha $, this formula describes the
asymptotic behavior of the VEV at the early stages of the cosmological
evolution corresponding to $t\rightarrow -\infty $. In this limit the
topological part behaves as $\exp [-(D+1)t/\alpha ]$ and, hence, it
dominates the part corresponding to the uncompactified dS spacetime which is
time independent. In particular, the total energy density is negative.

In the opposite limit of small values for the ratio $\eta /L_{p+1}$ we
introduce in the formulae for the VEV of the energy-momentum tensor an
integration variable $y=L_{p+1}x$ and expand the integrants over $\eta
/L_{p+1}$. For real values of the parameter $\nu $, for the energy density
to the leading order we find%
\begin{eqnarray}
\Delta _{p+1}\langle T_{0}^{0}\rangle _{p,q} &\approx &\frac{2^{\nu }D\left[
D/2-\nu +2\xi \left( 2\nu -D-1\right) \right] }{(2\pi
)^{(p+3)/2}L_{p+1}^{1-q}V_{q-1}\alpha ^{D+1}}\Gamma (\nu )  \notag \\
&&\times \left( \frac{\eta }{L_{p+1}}\right) ^{D-2\nu }\sum_{n=1}^{\infty
}\sum_{\mathbf{n}_{q-1}=-\infty }^{+\infty }\frac{f_{(p+1)/2-\nu
}(nL_{p+1}k_{\mathbf{n}_{q-1}})}{n^{(p+1)/2-\nu }}.  \label{T00smallEta}
\end{eqnarray}%
In particular, this energy density is positive for a minimally coupled
scalar field and for a conformally coupled massive scalar field. Note that
for a conformally coupled massless scalar the coefficient in (\ref%
{T00smallEta}) vanishes. For the vacuum stresses the second term on the
right of formula (\ref{DelTii}) is suppressed with respect to the first term
given by (\ref{A}) by the factor $(\eta /L_{p+1})^{2}$ for $i=1,\ldots ,p+1$%
, and by the factor $(\eta k_{i})^{2}$ for $i=p+2,\ldots ,D$. As a result,
to the leading order we have the relation (no summation over $i$)
\begin{equation}
\Delta _{p+1}\langle T_{i}^{i}\rangle _{p,q}\approx \frac{2\nu }{D}\Delta
_{p+1}\langle T_{0}^{0}\rangle _{p,q},\;\eta /L_{p+1}\ll 1,
\label{TiismallEta}
\end{equation}%
between the energy density and stresses, $i=1,\ldots ,D$. The coefficient in
this relation does not depend on $p$ and, hence, it takes place for the
total topological part of the VEV as well. Hence, in the limit under
consideration the topological parts in the vacuum stresses are isotropic and
correspond to the gravitational source with barotropic equation of state.
Note that this limit corresponds to late times in terms of synchronous time
coordinate $t$, $(\alpha /L_{p+1})e^{-t/\alpha }\ll 1$, and the topological
part in the VEV is suppressed by the factor $\exp [-(D-2\nu )t/\alpha ]$.
For a conformally coupled massless scalar field the coefficient of the
leading term vanishes and the topological parts are suppressed by the factor
$\exp [-(D+1)t/\alpha ]$. As the uncompactified dS part is constant, it
dominates the topological part at the late stages of the cosmological
evolution.

For small values of the ratio $\eta /L_{p+1}$ and for purely imaginary $\nu $%
, in the way similar to that used for the case of the field square we can
see that the energy density behaves like%
\begin{equation}
\Delta _{p+1}\langle T_{0}^{0}\rangle _{p,q}\approx \frac{4De^{-Dt/\alpha
}BB_{D}}{(2\pi )^{(p+3)/2}\alpha L_{p+1}^{p+1}V_{q-1}}\sin [2|\nu |t/\alpha
+2|\nu |\ln (L_{p+1}/\alpha )+\phi _{0}+\phi _{1}],  \label{T00ImEta}
\end{equation}%
where the coefficient $B_{D}$ and the phase $\phi _{1}$ are defined by the
relation%
\begin{equation}
|\nu |(1/2-2\xi )+i\left[ D/4-(D+1)\xi \right] =B_{D}e^{i\phi _{1}}.
\label{DefBD}
\end{equation}%
In the same limit, the main contribution into the vacuum stresses comes from
the term $A$ in (\ref{A}) and one has (no summation over $i$)%
\begin{equation}
\Delta _{p+1}\langle T_{i}^{i}\rangle _{p,q}\approx \frac{8|\nu
|e^{-Dt/\alpha }BB_{D}}{(2\pi )^{(p+3)/2}\alpha L_{p+1}^{p+1}V_{q-1}}\cos
[2|\nu |t/\alpha +2|\nu |\ln (L_{p+1}/\alpha )+\phi _{0}+\phi _{1}].
\label{TiiImEta}
\end{equation}%
As we see, in the limit under consideration to the leading order the vacuum
stresses are isotropic.

\section{Twisted scalar field}

\label{sec:Twisted}

\begin{figure}[tbph]
\begin{center}
\begin{tabular}{cc}
\epsfig{figure=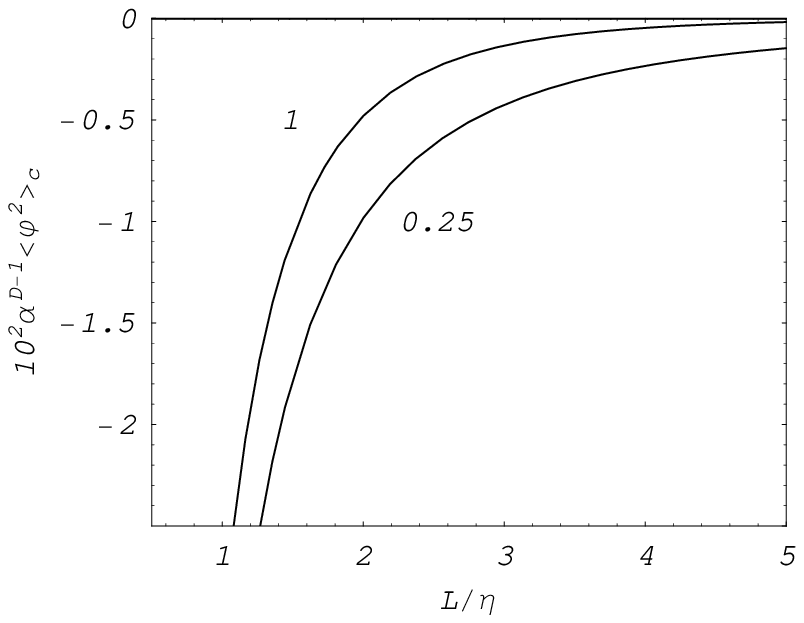,width=7.cm,height=6cm} & \quad %
\epsfig{figure=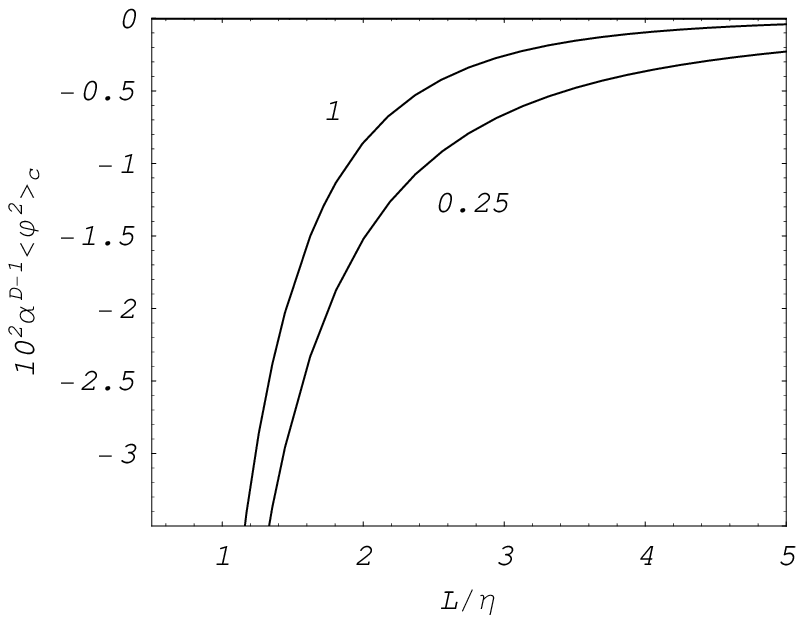,width=7.cm,height=6cm}%
\end{tabular}%
\end{center}
\caption{The topological part in the VEV of the field square in the case of
a conformally coupled twisted massive scalar ($\protect\xi =\protect\xi _{D}$%
) for $D=3$ dS spacetime with spatial topologies $\mathrm{R}^{2}\times
\mathrm{S}^{1}$ (left panel) and $(\mathrm{S}^{1})^{3}$ (right panel) as a
function of $L/\protect\eta =Le^{t/\protect\alpha }/\protect\alpha $. In the
second case we have taken the lengths for all compactified dimensions being
the same: $L_{1}=L_{2}=L_{3}\equiv L$. The numbers near the curves
correspond to the values of the parameter $m\protect\alpha $. }
\label{fig1}
\end{figure}

One of the characteristic features of field theory on backgrounds with
non-trivial topology is the appearance of topologically inequivalent field
configurations \cite{Isha78}. In this section we consider the case of a
twisted scalar field on background of dS spacetime with the spatial topology
$\mathrm{R}^{p}\times (\mathrm{S}^{1})^{q}$ assuming that the field obeys
the antiperiodicity condition (no summation over $l$)%
\begin{equation}
\varphi (t,\mathbf{z}_{p},\mathbf{z}_{q}+L_{l}\mathbf{e}_{l})=-\varphi (t,%
\mathbf{z}_{p},\mathbf{z}_{q}),  \label{AntiPer}
\end{equation}%
where $\mathbf{e}_{l}$ is the unit vector along the direction of the
coordinate $z^{l}$, $l=p+1,\ldots ,D$. The corresponding Wightman fucntion
and the VEVs of the field square and the energy-momentum tensor can be found
in the way similar to that for the field with periodicity conditions. The
eigenfunctions have the form given by (\ref{eigfuncD}), where now%
\begin{equation}
k_{l}=2\pi (n_{l}+1/2)/L_{l},\;n_{l}=0,\pm 1,\pm 2,\ldots ,\;l=p+1,\ldots ,D.
\label{nltwisted}
\end{equation}%
The positive frequency Wightman function is still given by formula (\ref%
{GxxD}). For the summation over $n_{p+1}$ we apply the Abel-Plana formula in
the form \cite{Most97,Saha07Gen}%
\begin{equation}
\sum_{n=0}^{\infty }f(n+1/2)=\int_{0}^{\infty }dx\,f(x)-i\int_{0}^{\infty
}dx\,\frac{f(ix)-f(-ix)}{e^{2\pi x}+1}.  \label{abel2}
\end{equation}%
Similar to (\ref{GxxD2}), for the correction to the Wightman function due to
the compactness of the $(p+1)$th spatial direction this leads to the result
\begin{eqnarray}
\Delta _{p+1}G_{p,q}^{+}(x,x^{\prime }) &=&-\frac{2\alpha ^{1-D}(\eta \eta
^{\prime })^{D/2}}{(2\pi )^{p+1}V_{q-1}}\int d\mathbf{k}_{p}\,e^{i\mathbf{k}%
_{p}\cdot \Delta \mathbf{z}_{p}}\sum_{\mathbf{n}_{q-1}=-\infty }^{+\infty
}e^{i\mathbf{k}_{q-1}\cdot \Delta \mathbf{z}_{q-1}}  \notag \\
&&\times \int_{0}^{\infty }dx\,\frac{x\cosh (\sqrt{x^{2}+\mathbf{k}%
_{p}^{2}+k_{\mathbf{n}_{q-1}}^{2}}\Delta z^{p+1})}{\sqrt{x^{2}+\mathbf{k}%
_{p}^{2}+k_{\mathbf{n}_{q-1}}^{2}}(e^{L_{p+1}\sqrt{x^{2}+\mathbf{k}%
_{p}^{2}+k_{\mathbf{n}_{q-1}}^{2}}}+1)}  \notag \\
&&\times \left[ K_{\nu }(\eta x)I_{-\nu }(\eta ^{\prime }x)+I_{\nu }(\eta
x)K_{\nu }(\eta ^{\prime }x)\right] ,  \label{GxxD2tw}
\end{eqnarray}%
where now $\mathbf{k}_{q-1}=(\pi (2n_{p+2}+1)/L_{p+2},\ldots ,\pi
(2n_{D}+1)/L_{D})$, and
\begin{equation}
k_{\mathbf{n}_{q-1}}^{2}=\sum_{l=p+2}^{D}\left[ \pi (2n_{l}+1)/L_{l}\right]
^{2}.  \label{knqtw}
\end{equation}%
Taking the coincidence limit of the arguments, for the VEV of the field
square we find

\begin{eqnarray}
\Delta _{p+1}\langle \varphi ^{2}\rangle _{p,q} &=&\frac{4\alpha ^{1-D}\eta
^{D}}{(2\pi )^{(p+3)/2}V_{q-1}}\sum_{n=1}^{\infty }(-1)^{n}\sum_{\mathbf{n}%
_{q-1}=-\infty }^{+\infty }\int_{0}^{\infty }dx\,xK_{\nu }(x\eta )  \notag \\
&&\times \frac{I_{-\nu }(x\eta )+I_{\nu }(x\eta )}{(nL_{p+1})^{p-1}}%
f_{(p-1)/2}(nL_{p+1}\sqrt{x^{2}+k_{\mathbf{n}_{q-1}}^{2}}),
\label{DelPhi2tw}
\end{eqnarray}%
with the notations being the same as in (\ref{DelPhi2}). Note that in this
formula we can put $\sum_{\mathbf{n}_{q-1}=-\infty }^{+\infty }=2^{q-1}\sum_{%
\mathbf{n}_{q-1}=0}^{+\infty }$. In particular, for the topology $\mathrm{R}%
^{D-1}\times \mathrm{S}^{1}$ with a single compactified dimension of the
length $L_{D}=L$, considered in \cite{Saha07}, we have $\langle \varphi
^{2}\rangle _{c}=\Delta _{D}\langle \varphi ^{2}\rangle _{D-1,1}$ with the
topological part given by the formula%
\begin{eqnarray}
\langle \varphi ^{2}\rangle _{c} &=&\frac{4\alpha ^{1-D}}{(2\pi )^{D/2+1}}%
\sum_{n=1}^{\infty }(-1)^{n}\int_{0}^{\infty }dx\,x^{D-1}  \notag \\
&&\times \left[ I_{-\nu }(x)+I_{\nu }(x)\right] K_{\nu }(x)\frac{%
K_{D/2-1}(nLx/\eta )}{(nLx/\eta )^{D/2-1}}.  \label{phi2SingComp}
\end{eqnarray}%
In figure \ref{fig1} we have plotted the topological part in the VEV of the
field square in the case of a conformally coupled twisted massive scalar ($%
\xi =\xi _{D}$) for $D=3$ dS spacetime with spatial topologies $\mathrm{R}%
^{2}\times \mathrm{S}^{1}$ (left panel) and $(\mathrm{S}^{1})^{3}$ (right
panel) as a function of $L/\eta =Le^{t/\alpha }/\alpha $. In the second case
we have taken the lengths for all compactified dimensions being the same: $%
L_{1}=L_{2}=L_{3}\equiv L$. The numbers near the curves correspond to the
values of the parameter $m\alpha $. Note that we have presented conformally
non-trivial examples and the graphs are plotted by using the general formula
(\ref{DelPhi2tw}). For the case $m\alpha =1$ the parameter $\nu $ is pure
imaginary and in accordance with the asymptotic analysis given above the
behavior of the field square is oscillatory for large values of the ratio $%
L/\eta $. For the left panel in figure \ref{fig1} the first zero is for $%
L/\eta \approx 8.35$ and for the right panel $L/\eta \approx 9.57$.

\begin{figure}[tbph]
\begin{center}
\begin{tabular}{cc}
\epsfig{figure=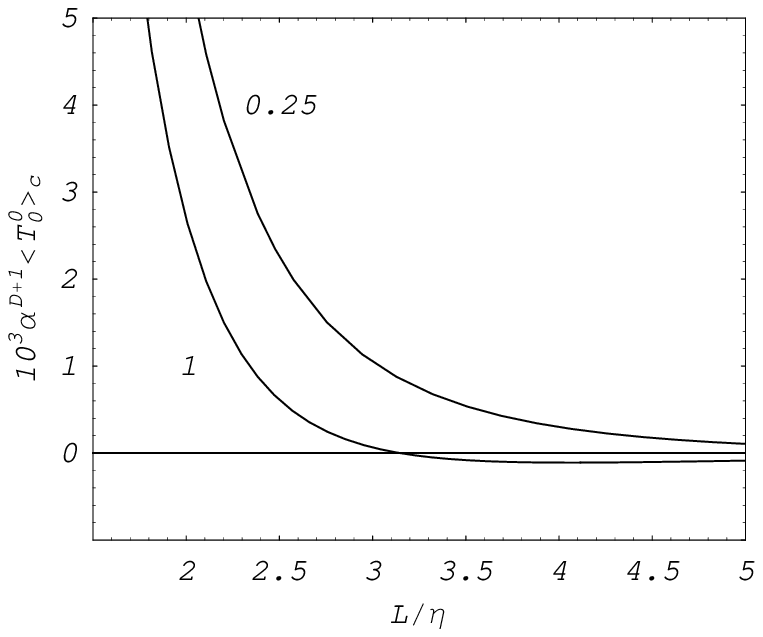,width=7.cm,height=6cm} & \quad %
\epsfig{figure=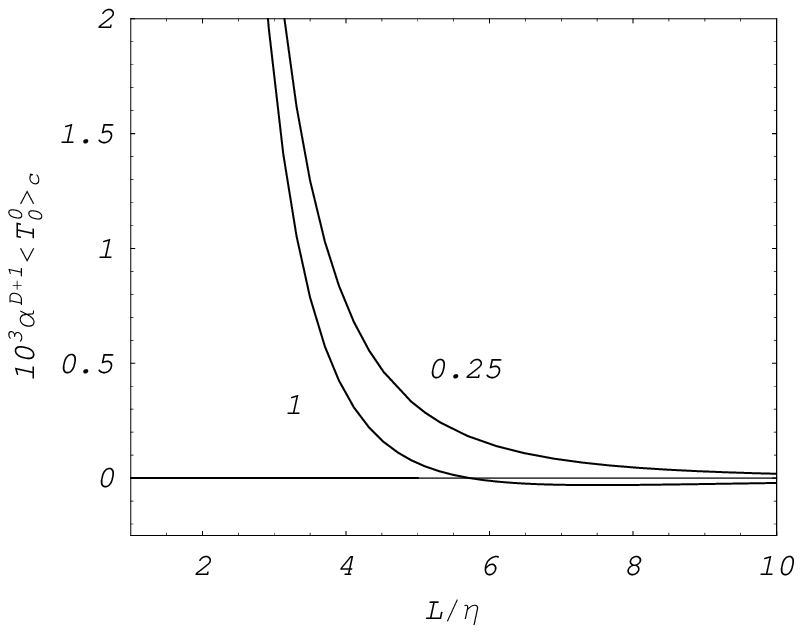,width=7.cm,height=6cm}%
\end{tabular}%
\end{center}
\caption{The same as in figure \protect\ref{fig1} for the topological part
of the energy density. }
\label{fig2}
\end{figure}

In the case of a twisted scalar field the formulae for the VEV of the
energy-momentum tensor are obtained from formulae for the untwisted field
given in the previous section (formulae (\ref{DelT00}), (\ref{DelTii})) with
$k_{\mathbf{n}_{q-1}}^{2}$ from (\ref{knqtw}) and by making the replacement%
\begin{equation}
\sum_{n=1}^{\infty }\rightarrow \sum_{n=1}^{\infty }(-1)^{n},\
\label{SumRepl}
\end{equation}%
and $k_{i}=2\pi (n_{i}+1/2)/L_{i}$ in expression (\ref{fp+1}) for $%
f^{(i)}(y) $, $i=p+2,\ldots ,D$. In figure \ref{fig2} the topological part
in the VEV of the energy density is plotted versus $L/\eta $ for a a
conformally coupled twisted massive scalar in $D=3$ dS spacetime with
spatial topologies $\mathrm{R}^{2}\times \mathrm{S}^{1}$ (left panel) and $(%
\mathrm{S}^{1})^{3}$ (right panel). In the latter case the lengths of
compactified dimensions are the same. As in figure \ref{fig1}, the numbers
near the curves are the values of the parameter $m\alpha $. For $m\alpha =1$
the behavior of the energy density for large values $L/\eta $ correspond to
damping oscillations. In the case $m\alpha =0.25$ (the parameter $\nu $ is
real) for the example on the left panel the topological part of the energy
density vanishes for $L/\eta \approx 9.2$, takes the minimum value $\langle
T_{0}^{0}\rangle _{c}\approx -3.1\cdot 10^{-6}/\alpha ^{4}$ for $L/\eta
\approx 12.9$ and then monotonically goes to zero. For the example on the
right panel with $m\alpha =0.25$ the energy density vanishes for $L/\eta
\approx 45$, takes the minimum value $\langle T_{0}^{0}\rangle _{c}\approx
-1.1\cdot 10^{-8}/\alpha ^{4}$ for $L/\eta \approx 64.4$ and then
monotonically goes to zero. For a conformally coupled massless scalar field
in the case of topology $(\mathrm{S}^{1})^{3}$ one has $\langle
T_{0}^{0}\rangle _{c}=0.1957(\eta /\alpha L)^{4}$. Note that in the case of
topology $\mathrm{R}^{D-1}\times \mathrm{S}^{1}$ for a conformally coupled
massless scalar we have the formulae (no summation over $l$)%
\begin{eqnarray}
\langle T_{l}^{l}\rangle _{c} &=&\frac{1-2^{-D}}{\pi ^{(D+1)/2}}\left( \frac{%
\eta }{\alpha L}\right) ^{D+1}\zeta _{\mathrm{R}}(D+1)\Gamma \left( \frac{D+1%
}{2}\right) ,  \label{TllConfTwS1} \\
\langle T_{D}^{D}\rangle _{c} &=&-D\langle T_{0}^{0}\rangle _{c},\;\xi =\xi
_{D},\;m=0,  \label{T00ConfTwS1}
\end{eqnarray}%
where $l=0,1,\ldots ,D-1$, and $\zeta _{\mathrm{R}}(x)$ is the Riemann zeta
function. The corresponding energy density is positive.

\section{Conclusion}

\label{sec:Conc}

In topologically non-trivial spaces the periodicity conditions imposed on
possible field configurations change the spectrum of the vacuum fluctuations
and lead to the Casimir-type contributions to the VEVs of physical
observables. Motivated by the fact that dS spacetime naturally arise in a
number of contexts, in the present paper we consider the quantum vacuum
effects for a massive scalar field with general curvature coupling in $(D+1)$%
-dimensional dS spacetime having the spatial topology $\mathrm{R}^{p}\times (%
\mathrm{S}^{1})^{q}$. Both cases of the periodicity and antiperiodicity
conditions along the compactified dimensions are discussed. As a first step
for the investigation of vacuum densities we evaluate the positive frequency
Wightman function. This function gives comprehensive insight into vacuum
fluctuations and determines the response of a particle detector of the
Unruh-DeWitt type. Applying the Abel-Plana formula to the corresponding
mode-sum, we have derived a recurrence relation which presents the Wightman
function for the $\mathrm{dS}_{D+1}$ with topology $\mathrm{R}^{p}\times (%
\mathrm{S}^{1})^{q}$ in the form of the sum of the Wightman function for the
topology $\mathrm{R}^{p+1}\times (\mathrm{S}^{1})^{q-1}$ and the additional
part $\Delta _{p+1}G_{p,q}^{+}$ induced by the compactness of the $(p+1)$th
spatial dimension. The latter is given by formula (\ref{GxxD2}) for a scalar
field with periodicity conditions and by formula (\ref{GxxD2tw}) for a
twisted scalar field. The repeated application of formula (\ref{G1decomp})
allows us to present the Wightman function as the sum of the uncompactified
dS and topological parts, formula (\ref{DeltaGtop}). As the toroidal
compactification does not change the local geometry, by this way the
renormalization of the bilinear field products in the coincidence limit is
reduced to that for uncompactifeid $\mathrm{dS}_{D+1}$.

Further, taking the coincidence limit in the formulae for the Wightman
function and its derivatives, we evaluate the VEVs of the field square and
the energy-momentum tensor. For a scalar field with periodic conditions the
corresponding topological parts are given by formula (\ref{DelPhi2}) for the
field square and by formulae (\ref{DelT00}) and (\ref{DelTii}) for the
energy density and vacuum stresses respectively. The trace anomaly is
contained in the uncompactified dS part only and the topological part
satisfies the standard trace relation (\ref{tracerel}). In particular, this
part is traceless for a conformally coupled massless scalar. In this case
the problem under consideration is conformally related to the corresponding
problem in $(D+1)$-dimensional Minkowski spacetime with the spatial topology
$\mathrm{R}^{p}\times (\mathrm{S}^{1})^{q}$ and the topological parts in the
VEVs are related by the formulae $\langle \varphi ^{2}\rangle _{c}=(\eta
/\alpha )^{D-1}\langle \varphi ^{2}\rangle _{c}^{\mathrm{(M)}}$ and $\langle
T_{i}^{k}\rangle _{c}=(\eta /\alpha )^{D+1}\langle T_{i}^{k}\rangle _{c}^{%
\mathrm{(M)}}$. Note that for a conformally coupled massless scalar the
topological part in the energy density is always negative and is equal to
the vacuum stresses along the uncompactified dimensions.

For the general case of the curvature coupling, in the limit $L_{p+1}/\eta
\ll 1$ the leading terms in the asymptotic expansion of the VEVs coincide
with the corresponding expressions for a conformally coupled massless field.
In particular, this limit corresponds to the early stages of the
cosmological expansion, $t\rightarrow -\infty $, and the topological parts
behave like $e^{-(D-1)t/\alpha }$ for the field square and like $%
e^{-(D+1)t/\alpha }$ for the energy-momentum tensor. Taking into account
that the uncompactified dS part is time independent, from here we conclude
that in the early stages of the cosmological evolution the topological part
dominates in the VEVs. In the opposite asymptotic limit corresponding to $%
\eta /L_{p+1}\ll 1$, the behavior of the topological parts depends on the
value of the parameter $\nu $. For real values of this parameter the leading
terms in the corresponding asymptotic expansions are given by formulae (\ref%
{DelPhi2Mets}) and (\ref{T00smallEta}) for the field square and the
energy-momentum tensor respectively. The corresponding vacuum stresses are
isotropic and the topological part of the energy-momentum tensor corresponds
to the gravitational source of the barotropic type with the equation of
state parameter equal to $-2\nu /D$. In the limit under consideration the
topological part in the energy density is positive for a minimally coupled
scalar field and for a conformally coupled massive scalar field. In
particular, this limit corresponds to the late stages of the cosmological
evolution, $t\rightarrow +\infty $, and the topological parts of the VEVs
are suppressed by the factor $e^{-(D-2\nu )t/\alpha }$ for both the field
square and the energy-momentum tensor. For a conformally coupled massless
field the coefficient of the leading term in the asymptotic expansion
vanishes and the topological part is suppressed by the factor $%
e^{-(D+1)t/\alpha }$. In the limit $\eta /L_{p+1}\ll 1$ and for pure
imaginary values of the parameter $\nu $ the asymptotic behavior of the
topological parts in the VEVs of the field square and the energy-momentum
tensor is described by formulae (\ref{DelPhi2MetsIm1}), (\ref{T00ImEta}), (%
\ref{TiiImEta}). These formulae present the leading term in the asymptotic
expansion of the topological parts at late stages of the cosmological
evolution. In this limit the topological terms oscillate with the amplitude
going to the zero as $e^{-Dt/\alpha }$ for $t\rightarrow +\infty $. The
phases of the oscillations for the energy density and vacuum stresses are
shifted by $\pi /2$.

In section \ref{sec:Twisted} we have considered the case of a scalar field
with antiperiodicity conditions along the compactified directions. The
Wightman fucntion and the VEVs of the field square and the energy-momentum
tensor are evaluated in the way similar to that for the field with
periodicity conditions. The corresponding formulae are obtained from the
formulae for the untwisted field with $k_{\mathbf{n}_{q-1}}^{2}$ defined by
Eq. (\ref{knqtw}) and by making the replacement (\ref{SumRepl}). In this
case we have also presented the graphs of the topological parts in the VEVs
of the field square and the energy-momentum tensor for $\mathrm{dS}_{4}$
with the spatial topologies $\mathrm{R}^{2}\times \mathrm{S}^{1}$ and $(%
\mathrm{S}^{1})^{3}$.

\section*{Acknowledgments}

AAS would like to acknowledge the hospitality of the INFN Laboratori
Nazionali di Frascati, Frascati, Italy. The work of AAS was supported in
part by the Armenian Ministry of Education and Science Grant. The work of SB
has been supported in part by the European Community Human Potential Program
under contract MRTN-CT-2004-005104 "Constituents, fundamental forces and
symmetries of the Universe" and by INTAS under contract 05-7928.

\end{document}